\documentclass[english,conference]{IEEEtran}
\usepackage{mathrsfs}
\usepackage{amsfonts}
\usepackage{graphicx,cite,epsfig,amssymb,amsmath}
\usepackage{epstopdf}
\usepackage{color,xcolor}
\usepackage{pifont}
\usepackage{stmaryrd}
\usepackage{setspace}
\usepackage{subfigure}
\usepackage{cite}
\usepackage{array}\usepackage{float}\usepackage{multirow}
\usepackage{algorithmic,algorithm}

\floatstyle{ruled}
\newfloat{algorithm}{tbp}{loa}
\providecommand{\algorithmname}{Algorithm}
\floatname{algorithm}{\protect\algorithmname}

\begin{document}
\title{An Edge-Computing Based Architecture for Mobile Augmented Reality}
\author{Jinke Ren, Yinghui He, Guan Huang, Guanding Yu, Yunlong Cai, Zhaoyang Zhang\\
\IEEEauthorblockA{Department of Information Science and Electronic Engineering, Zhejiang University, Hangzhou 310027, China}
}
\maketitle
\begin{abstract}
In order to mitigate the long processing delay and high energy consumption of mobile augmented reality (AR) applications, mobile edge computing (MEC) has been recently proposed and is envisioned as a promising means to deliver better quality of experience (QoE) for AR consumers. In this article, we first present a comprehensive AR overview, including the indispensable components of general AR applications, fashionable AR devices, and several existing techniques for overcoming the thorny latency and energy consumption problems. Then, we propose a novel hierarchical computation architecture by inserting an edge layer between the conventional user layer and cloud layer. Based on the proposed architecture, we further develop an innovated operation mechanism to improve the performance of mobile AR applications. Three key technologies are also discussed to further assist the proposed AR architecture. Simulation results are finally provided to verify that our proposals can significantly improve the latency and energy performance as compared against existing baseline schemes.
\end{abstract}
\IEEEpeerreviewmaketitle
\section{Introduction}
As a revolutionary innovation, mobile augmented reality (AR) has become a miraculous technology in the past few years. By combining the computer-generated and sensor-extracted elements with the real objects and enabling real-time 3D interaction between mobile users and physical surroundings at fingertips, mobile AR is envisioned as a new paradigm to submerge mobile subscribers in a fabulous mixed-reality world \cite{AR_Survey, VM_cloudlet}. In view of the great business opportunities involved in mobile AR, many leading companies have continuously scrambled to design and promote their own mobile AR applications and products, such as Google Glass, Microsoft HoloLens, and Recon Jet.

However, despite the fully immersive user experience, mobile AR still faces with many technical challenges that hinder its widespread commercial use. Due to the hardware limitation, running sophisticated AR algorithms at mobile devices generally causes long processing delay and high energy consumption, which are the main hurdles for implementation. To handle these issues, many cutting-edge technologies, such as advanced hardware structure, acceptable approximate computing, and partial video frame updating  have been developed to improve the mobile AR performance. Nevertheless, these methods still cannot catch up with the rapid growth of user demand while pointlessly increasing the cost of mobile devices. Hence, novel techniques are urgently expected to tackle the above challenges.

To overcome the computational resource shortage of mobile devices, another major innovation in the past decade is \textit{mobile cloud computing} (MCC), which allows users to offload computation-intensive tasks to a number of powerful cloud servers deployed at the remote cloud platform for processing \cite{Cloud_computing}. However, with the strict delay requirement of mobile AR applications, MCC suffers from extra propagation delay due to the long physical distance between mobile devices and cloud servers \cite{MCC_magazine}. To this end, an improved technique, which is referred to as \textit{mobile edge computing} (MEC), has been recently proposed by the European Telecommunication Standards Institute (ETSI) and is also considered by the Third Generation Partnership Project (3GPP) in their future standards. By distributing the conventional centralized cloud computing resources to the edge of mobile networks, MEC offers an adjacent computing environment for mobile subscribers and provides a variety of benefits, including ultra-low latency, real-time access, and location-aware services, etc.\cite{MEC_magazine_paradigm, MEC_Survey, MEC_magazine_smartcity}.

On the other hand, by taking advantages of the proximity to mobile users in MEC and the abundant computation capacity in MCC, effective collaboration between cloud and edge computing can further improve the system performance. Several hierarchical edge-to-cloud architectures have been proposed to collaborate the computation capacity of edge servers and cloud servers\cite{Cloud_Edge1, Cloud_Edge2}. However, they are not specifically designed for AR applications. Motivated by this, in this work, we propose a novel hierarchical computation architecture to deal with the long processing delay and high energy consumption of mobile AR applications. The newly-designed architecture is composed of three layers: the user layer, the edge layer, and the cloud layer. By integrating the communication, computation, and control modules in the edge layer while seamlessly collaborating the edge and cloud computing capacities, several critical components of AR applications can be intelligently offloaded to both edge and cloud servers for further processing. In accordance with this architecture, an advanced operation mechanism is developed by embedding the MEC functions into the AR process. Moreover, to support the proposed AR framework, we further discuss three key technologies, i.e., the joint communication and computation resource allocation, the collaborative cloud and edge computing, and the content-based image retrieval from the viewpoints of energy efficiency and delay optimization. The main merit of the proposed MEC-based AR framework is that it merges the advantages of cloud computing and edge computing to enormously improve the quality of experience (QoE) for AR applications.

In what follows, we first provide a comprehensive AR overview, including the indispensable components of general AR applications, fashionable AR devices, and existing technologies for alleviating the inherent long processing delay and high energy consumption problems. To improve the system performance, our key contributions are the novel hierarchical computation architecture and the related AR operation mechanism as well as the three key technologies, which can significantly facilitate the implementation of the MEC-based AR framework. Finally, simulation results demonstrate that the proposed AR framework can evidently improve the energy and delay performance, as compared against some baseline schemes. Our design in this article has the great potential to provide an essential reference for the future development of mobile AR technology.

\section{Overview of Augmented Reality}
In this section, we present an overview of AR, including the specific components of general AR applications, the compelling mobile AR devices, and the incumbent AR technologies, from the viewpoints of energy efficiency and latency optimization.
\subsection{AR Application Components}
To combine colorful computer-generated and sensor-extracted data with the physical reality, different AR applications possess diverse modules for specific processing purposes. However, five computation components, i.e., the $\emph{video source}$, $\emph{tracker}$, $\emph{mapper}$, \emph{object recognizer}, and $\emph{renderer}$, are indispensable for almost all AR applications, and they play the critical roles in general AR architectures \cite{AR_cloudlet_1, AR_cloudlet_2}.
\begin{figure}[!htp]
	\centering
	\includegraphics[width=0.5\textwidth]{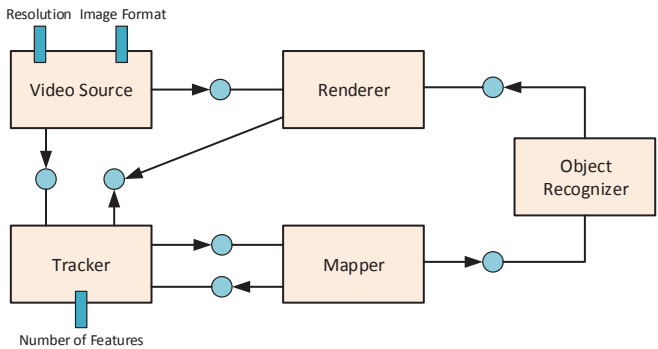}
	\caption{Main components of general AR applications.}	
	\label{Main Components of an AR Application}
\end{figure}

As illustrated in Fig. \ref{Main Components of an AR Application}, the aforementioned five components collaborate closely to accomplish an integrated AR process. First, the video source fetches the raw videos from mobile cameras and clips these videos into frames with specific image format, such as JPEG and PNG. Next, the video frames are delivered to the tracker to determine the user's position with respect to the physical surroundings. Given the tracking results, virtual coordinate of the environment can be established by the mapper. Then, the internal objects in video frames are identified by the object recognizer with robust features. Afterwards, the augmented information of the identified object can be accurately retrieved from local memory (or cloud database) and properly mixed with the original videos by the renderer. The results are finally displayed on the screens of AR devices so that the subscribers are able to enjoy a magic interactive experience with the physical reality at the fingertips.
\subsection{Fashionable AR Devices}
A variety of AR devices with different operation modes have been recently designed with multifarious functions. Table \uppercase\expandafter{\romannumeral1} summarizes several well-known AR devices developed by major companies. Among them, Google Glass, Microsoft HoloLens, and Osterhout Design Group (ODG) R-9 operate independently, whereas Meta 2 and Recon Jet need to rely on external equipments such as personal computers (PCs) or smart phones.

The Google Glass is the first-appeared AR glass, which has a stylish appearance and provides users with requested information, such as calendar, weather, and message via natural language voice commands. Comparatively, Microsoft HoloLens enables users to engage with digital content and interact with holograms of the surroundings for supplying subscribers with a mixed-reality experience. This function is extended in ODG R-9 by combining with some advanced modules, such as Bluetooth 5.0, built-in GPS, and six degrees of freedom tracking. However, these devices all suffer from high energy consumption and cannot work for a long time. On the other hand, by connecting powerful computers, Meta 2 allows users to intuitively touch, grab, and move the computer-generated digital objects as real ones, which inevitably leads to its non-portability for outdoor use. Particularly, Recon Jet is extensively used in sport scenes and can exhibit the real-time motion information of mobile users.
\begin{table*}
	\centering
    \footnotesize
	\caption{Major AR Devices}
    \vspace{-1em}
	\label{table1}
    \begin{tabular}{p{1.5cm}p{1.5cm}p{1.5cm}p{1.5cm}p{3.1cm}p{2.8cm}p{1cm}}
  	\hline
  	\hline
  	Product&Operation Mode  &CPU &GPU &Latency &Battery Autonomy &Weight \\
  	\hline
	\hline
    Google Glass  &Independent &1 GHz &300 MHz &700 ms for user interface (UI) response &$<$ 1 h when recording videos &42 g  \\
    \hline
  	Microsoft HoloLens &Independent &1.04 GHz 	 &HoloLens Graphics  &Few seconds for gesture recognition    &2-3 h for typical use   &579 g \\
  	\hline
	ODG R-9 & Independent &2.45 GHz &710 MHz &Not available &24 h for browsing and casual use &184 g\\
    \hline
  	Meta 2  &Dependent &3.4 GHz &1127 MHz &80 ms for gesture recognition &Powered by PC &500 g \\
    \hline
	Recon Jet &Dependent &1 GHz &PowerVR SGX540 &40 s for loading navigation map &4 h for typical use &85 g  \\
	\hline
	\hline
	\end{tabular}
\end{table*}

Although the above AR devices can provide fully immersive experience for mobile subscribers, they generally suffer from long processing delay and high energy consumption, which are the most-critical performance metrics for AR applications. To overcome these drawbacks, several existing technologies have been recently developed, which will be discussed in the following.
\subsection{Existing Technologies}
Presently, there are mainly three kinds of technologies to reduce the processing delay and energy consumption of AR applications, as summarized in the following \cite{AR_Survey}.
\begin{itemize}
    \item   \textbf{Advanced hardware structure:}
            It is hard for traditional battery to sustainably support the complex operations of AR applications. As a result, advanced hardware structures such as multi-core CPU with low frequency and voltage can be utilized to replace the single-core ones. Furthermore, the dynamic voltage and frequency scaling (DVFS) technology can be adopted to save energy with respect to specific requirements.
    \item   \textbf{Acceptable approximate computing:}
            Consistently performing accurate calculation in AR applications usually consumes much energy and results in long processing delay. Therefore, approximate computing for AR tasks can be adopted according to detailed computational accuracy requirements. For example, low-precision tasks, such as location sensing, can be approximately processed to balance computational accuracy and energy (delay) efficiency.
    \item   \textbf{Partial video frame updating:}
            The frequent movement of mobile AR devices would produce a large volume of video data, which is adverse for real-time processing. Therefore, partial video frame updating can be exploited to avoid redundant computing by processing the newly-generated data only.
\end{itemize}

Based on the above mentioned technologies, the end-to-end latency and energy consumption of AR applications can be improved to some extent. However, since the computation-intensive tasks, e.g., tracking, mapping, and object recognition, are still remained for local computing, the end-to-end latency and energy consumption cannot be further reduced and the anticipated user demand is still far from reaching.
To overcome the resource shortage of mobile devices, MEC has been recently proposed and is regarded as a new paradigm to deliver better experience for mobile users by offloading computation-intensive tasks to the servers deployed at the edge of mobile networks \cite{MEC_You}. Motivated by this, we present a novel computation framework for embedding MEC into AR applications in the following section.
\section{MEC-Based AR Framework}
In this section, we first present a novel hierarchical computation architecture for multi-user AR systems and then introduce a detailed mechanism to support AR applications.
\subsection{Hierarchical Computation Architecture}
\begin{figure}[!htp]
	\centering
	\includegraphics[width=0.5\textwidth]{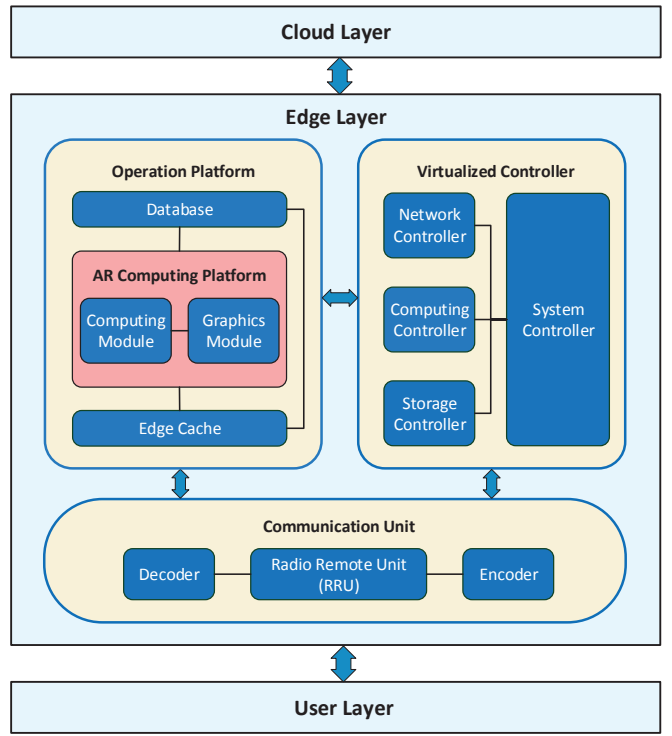}
	\caption{Hierarchical Computation Architecture.}	
	\label{Hierarchical Computation Architecture}
\end{figure}

Considering a multi-user AR scenario where several AR applications need to be executed simultaneously, we propose a hierarchical architecture composed of three layers, namely, the \textit{user layer}, the \textit{edge layer}, and the \textit{cloud layer}. As illustrated in Fig. \ref{Hierarchical Computation Architecture}, multiple AR devices are located in the user layer and are connected to the edge layer through wireless links. The cloud layer mainly consists of an enormous database for data storage and abundant computational resources for data processing. These two layers are quite similar to the existing AR architectures and thus we do not devote special attention henceforth.

The uniqueness of our proposed architecture is that an edge layer is inserted between the conventional cloud layer and user layer. The edge layer can be equipped at the base stations (BSs) of cellular networks or the access points (APs) of WiFi networks. The merit of this design is that MEC can be utilized to improve the end-to-end latency and energy consumption performance of AR applications. Furthermore, cloud computing and edge computing can be potentially collaborated for further performance enhancement.

In the proposed hierarchical architecture, one major issue is how to design the function modules in the edge layer to seamlessly collaborate with the other two layers. To tackle this problem, we introduce three functionalized components, namely, $\emph{communication unit}$, $\emph{operation platform}$, and $\emph{virtualized controller}$ in the edge layer on the basis of the software defined network (SDN) technology. Note that the combination of the three modules also leads to the convergence of communication, computation, and control. In what follows, we will present a detailed description on these three modules.
\begin{itemize}
    \item   \textbf{Communication Unit:}
             The communication unit can be regarded as a ``bridge" that enables real-time data transmission between the edge layer and the other two layers. On one aspect, when multiple AR devices offload the computational tasks, such as video steams, to the BS, the inner radio remote unit (RRU) has the responsibility to successfully receive these data. Furthermore, these data are required to be delivered to the operation platform for further processing. On the other aspect, the computation results from operation platform and cloud layer should be multicasted to the corresponding users through the communication unit.
    \item   \textbf{Operation Platform:}
            The operation platform is the core of this architecture, which processes the offloaded AR tasks from mobile users. The original data from RRU is firstly stored in the edge cache, and will be delivered to the computing platform for further processing. The critical AR computing platform is composed of a computing module (resembling computer's CPU) and a graphics module (resembling computer's GPU). The former is utilized to process the computing-related tasks, such as tracking and mapping, while the latter is utilized to process the graphics-related tasks like object recognition. Since AR tasks usually require additional data, such as 3D-models and annotations of the recognized objects, we establish a small database at the edge layer for storing the object information that is frequently accessed. With this design, we do not have to invariably fetch the requested information from the remote cloud database, which can significantly reduce the end-to-end latency.

    \item   \textbf{Virtualized Controller:}
            The virtualized controller serves as the ``centralized coordinator" of the whole edge layer, which is divided into four specific components: the network controller, the computing controller, the storage controller, and the system controller. The network controller manages all network activities among three layers, such as network establishment and data transmission. Accordingly, the computing controller supervises the entire process in the operation platform while it optimally allocates the available computational resource to each AR task with specific requirements. Moreover, the inherent executive priorities and collaborative properties of AR tasks from different subscribers are also evaluated in the computing controller. In addition, the storage controller aims at properly managing the memory mechanism of the edge database for fast data searching and updating. Finally, the system controller monitors the behaviours of the above three controllers and coordinates them in a more efficient way.

\end{itemize}
\subsection{MEC-based AR Operation Mechanism}
Thus far, our discussions have focused on the physical modules of the proposed hierarchical architecture for AR applications. In the following, we further present some detailed operations to facilitate the AR implementation.
\begin{figure}[!htp]
	\centering
	\includegraphics[width=0.5\textwidth]{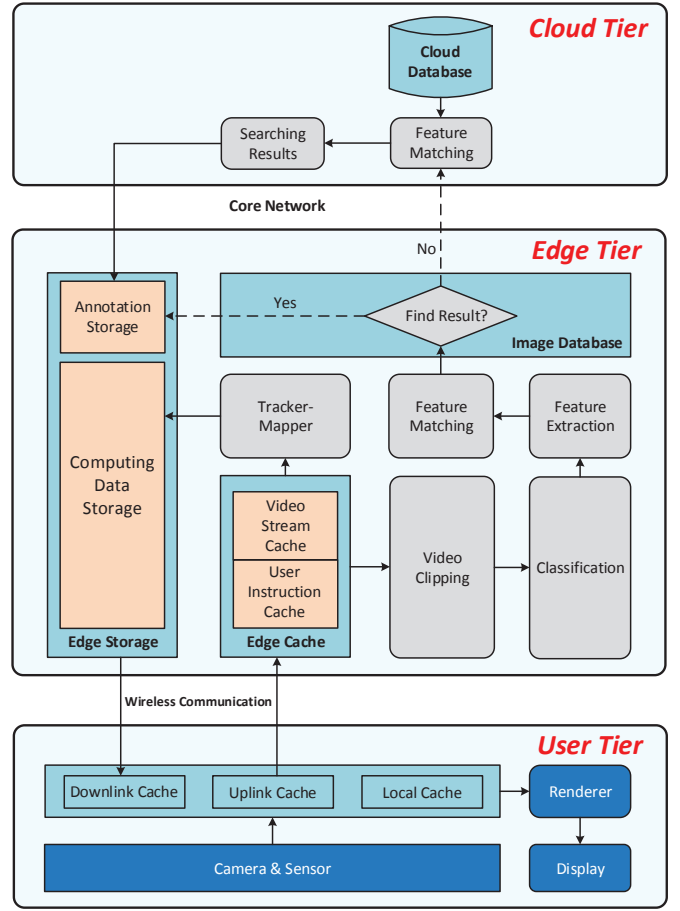}
	\caption{MEC-based AR operation mechanism.}	
	\label{Novel AR Executing Mechanism}
\end{figure}

As depicted in Fig. \ref{Novel AR Executing Mechanism}, the proposed MEC-based AR operation mechanism can be divided into three tiers, which are consistent with the hierarchical architecture, i.e., the \textit{user tier}, the \textit{edge tier}, and the \textit{cloud tier}. The main idea of this mechanism is that the video source and renderer of AR applications must be processed locally whereas the computation-intensive components, i.e., the tracker, mapper, and object recognition should be offloaded to the edge and cloud servers. The detailed operations in each tier are generalized into specific blocks and will be elaborated in the following.

The user tier is responsible for executing the local operations. Multiple AR devices simultaneously start with sensing real environment, producing raw videos, and capturing users' gestures via their cameras and sensors. Through identification analysis, this original information will be transformed into two categories: the video streams which contain the raw video data and the operation instructions which carry the specific requirements of mobile subscribers, such as object identification and sentiment analysis. Thereafter, this information will be further duplicated into two copies with one stored in the uplink cache for data transmission and the other stored in the local cache for subsequent processing. Consequently, the transceivers of mobile AR devices will transmit the data in the uplink cache to the edge layer through wireless channels.

The edge tier plays a critical role in computing AR applications. Upon receiving the offloaded data from AR devices, the BS will first classify them into two kinds: the raw video streams and the user's operation instructions, and then deliver them to the edge cache for separate storage. Thereafter, these two kinds of data are delivered to the tracker-mapper module and the video clipping module, respectively. The tracker-mapper module then tracks users' locations and builds virtual coordinates to coincide with the real world via some algorithms, such as simultaneous localization and mapping (SLAM) and parallel tracking and mapping (PTAM). Meanwhile, the video clipping module slices one representative frame (or image) from each raw video for subsequent processing. It should be noted that some users may observe the same object and require the same information of this object. With this regard, classifiers can be leveraged to assort all frames from different videos into several categories based on the inherent information, such that each category contains those frames of the same object. This function can be implemented via some well-known machine learning algorithms, such as convolutional neural networks (CNN) and support vector machines (SVM). Afterwards, one typical image of each category is picked out and utilized to match with the standard images pre-stored in the edge database by some image retrieval methods. By this means, the inherent collaborative properties of AR applications can be fully utilized and redundant calculation can be effectively avoided. Then, if the matched standard image is found in the edge database, the corresponding annotation information will be fetched from the adjacent edge storage. Otherwise, the related frame will be offloaded to the cloud server for further searching.

The cloud tier contains a large cloud database for storing the additional data that are not cached in the edge tier due to its limited memory size. Once the retrieval operation in the edge tier is not realized, the corresponding image will be offloaded to the cloud tier for further searching. Owing to the adequate computational resource at the cloud server and the sufficient capacity of the cloud storage, the image retrieval will be generally achieved. Thereafter, the requested information will be transmitted back to the edge tier and combined with the virtual map that is reconstructed by the tracker-mapper module. Also, the total computation results are multicasted to the corresponding devices. Finally, these data are exhibited to the subscribers after mixing with the original videos via the local renderers.
\section{Key Implement Technologies}
In this section, we will develop three key technologies, i.e., the joint communication and computation resource allocation, the collaborative cloud and edge computing, and the content-based image retrieval to further improve the energy efficiency and reduce the processing latency.
\subsection{Joint Communication and Computation Resource Allocation}
In the proposed framework, the computational tasks of an AR application can be jointly processed by the local device, the edge server, and the cloud server. Meanwhile, the communication resource of the wireless network and the computational resources of the edge/cloud servers can be shared by all AR devices. Therefore, joint communication and computation resource allocation is essential to further improve the delay and energy performance \cite{Joint_resource_allocation}. Generally, two schemes in current AR scenarios should be taken into account, i.e., the centralized and distributed resource allocation schemes.

In centralized AR systems, both user priority and channel state information can be acquired in the edge layer. Therefore, the network controller and the computing controller can collaboratively design the resource allocation policy by optimizing a specific objective function, such as minimizing the end-to-end  latency under a prescribed resource utilization constraint and maximizing the mobile energy efficiency under an offloading latency constraint. Moreover, the network controller and the computing controller can monitor the specific requirements of different AR tasks in real-time and adaptively adjust the resource allocation to meet the diverse user demands.

However, in distributed AR systems, the aforementioned information can no longer be obtained by the edge layer, making it difficult to centrally control the allocation of available communication and computation resources. To tackle this problem, game-theoretical techniques can be exploited to develop distributed algorithms based on past network and channel information, whereas the obtained result is demonstrated to achieve the Nash equilibrium and can deliver satisfactory latency and energy consumption experiences for mobile AR subscribers.
\subsection{Collaborative  Cloud and Edge Computing}
On the one hand, traditional cloud servers are typically deployed at the remote cloud platform, suffering from long propagation delay when transmitting onerous AR tasks in the core network. On the other hand, although the adjacent MEC servers can be implemented at the nearby BS, their computation and storage capacities are usually limited. Therefore, by combining the advantages of the powerful computation capacity of cloud computing and the proximity to mobile subscribers of edge computing, collaborative cloud and edge computing is envisioned as a promising paradigm to achieve better AR performance.

In the proposed framework, due to the limited edge storage capacity, the additional information such as 3D models and annotations of those unpopular objects should be stored in the cloud database to alleviate the storage burden at the edge database. On the other hand, the edge database only need to store those frequently accessed information of the popular objects. This can be realized by storing the historical access record of each object's information in the edge layer and setting an appropriate threshold to judge which object is popular. By this means, massive object recognition operations can be avoided at the edge layer and the AR processing latency can be effectively reduced. Moreover, when several edge servers are simultaneously served by the same cloud server, it is important to properly allocate the available cloud computational resource to each edge server based on their workload and local computation capacities. The detailed resource allocation policy can be derived by solving a specific optimization problem through some convex and non-convex optimization tools. For instance, more cloud computational resource need to be allocated to assist those edge servers with less computation capacities and heavier workload to balance the uneven resource and workload distribution over different edge servers.
\subsection{Content-based Image Retrieval}
Object recognizer is a critical component of AR applications, which usually consumes a large processing delay, especially in the image retrieval procedure. Therefore, advanced content-based image retrieval technology should be developed to accelerate the searching speed. Motivated by the work in \cite{MVS}, we present a typical pipeline for image retrieval within the proposed mechanism, which consists of three steps: feature extraction, feature matching, and geometric verification.

\textbf{1) Feature extraction:}
Once an image is input into the graphics module of the edge layer, the feature extraction algorithm will immediately search its inherent salient interest points, which are used to estimate the similarity between this image and the standard images pre-stored in the edge and cloud databases. Various robust feature descriptors can be applied to obtain typical features, such as scale invariant feature transform and speeded up robust features \cite{Feature_extraction}.

\textbf{2) Feature matching:}
Current pairwise feature matching algorithm generally consumes a lot of time since it directly matches the input image with all standard images. To cope with this issue,  we pre-construct a data structure to store the features of all standard images with particular indices. Then, by comparing the extracted features with those in the data structure, a shortlist of candidate images can be efficiently filtered. Thereafter, the slow pairwise feature matching method can be applied between the input image and the candidate images only, and a best-matched image can be eventually picked out from the shortlist. In this way, the image retrieval delay can be substantially reduced.

\textbf{3) Geometric verification:}
After obtaining the best-matched standard image, further examination is required to confirm whether the matching result is correct. Geometric verification is commonly utilized to test whether the input image and the best-matched standard image are similar with only geometric and photometric distortions. If the verification result is correct, the matching relation will be established. Otherwise, the input image and its features will be offloaded to the cloud server for further searching.

It shall be noted that with the above three steps, the computation and storage resources of the cloud server and edge server can be jointly utilized to improve the accuracy of object recognition. Consequently, the end-to-end latency and energy consumption can be effectively reduced.
\section{Performance Evaluation}
In this section, we present simulation results to manifest the latency and energy performance enhancement of our proposals against two benchmark schemes. We consider a scenario where multiple mobile devices in the coverage of the same BS execute AR applications simultaneously. The edge layer is implemented at the central BS with a radius of 200 m. The cloud layer is deployed in a remote cloud platform. The BS connects the cloud platform through a backhaul link, whose uplink and downlink transmission capacities are both set to be 200 Mbps. The edge and cloud computing capacities are set to be $\text{5} \times \text{10}^{\text{10}}$ CPU cycle/s and $\text{3} \times \text{10}^{\text{11}}$ CPU cycle/s, respectively \cite{MEC_You}. Each AR device is randomly distributed according to Poisson distribution within the BS coverage, adopting the TDMA channel access with a bandwidth of 15 MHz. The local computation capacity of each device follows the uniform distribution between $[5 \times 10^{8}, 2 \times 10^{9}]$  CPU cycle/s. The channel gains between mobile devices and the BS are generated according to independent and identically distributed (i.i.d.) Rayleigh random variables with unit variance. Since the cloud computing components are first offloaded to the edge layer through wireless links and then being transmitted to the cloud server through the backhaul link equipped with high transmission bandwidth, the distance between mobile devices and the cloud server could be ignored in this paper.
Other major parameters of AR applications can be found in \cite{MEC_AR_Letter}.

In our test, we compare the performance of our proposals with two benchmark schemes: the local computing scheme where the tracker and mapper modules are remained for local processing while the object recognizer is offloaded for cloud computing, and the cloud computing scheme where the tracker, mapper, and object recognizer are all offloaded to the remote cloud platform. Meanwhile, we name our proposed scheme as the edge computing scheme.

\begin{figure}
	\centering
	\subfigure[End-to-end latency of three schemes.]{
		\includegraphics[width=0.75\columnwidth]{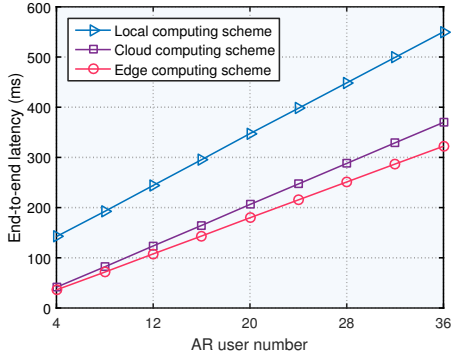} \label{Per-user delay}

	}
	\subfigure[Energy consumption of three schemes.]{
		\includegraphics[width=0.75\columnwidth]{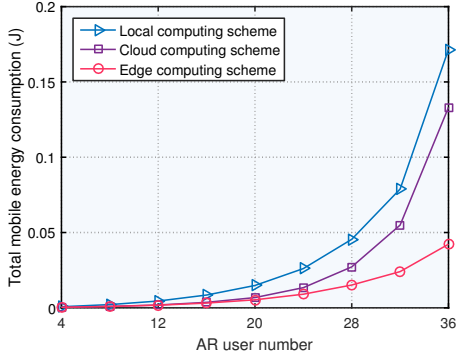} \label{Total-user energy}
	}
 \subfigure[Total mobile energy consumption under different delay tolerances.]{
		\includegraphics[width=0.75\columnwidth]{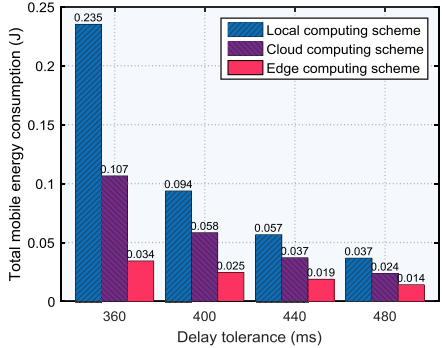} \label{Total_Energy_Delay_tolerance}
	}
	\caption{Simulation results.}\label{Simulation results}
\end{figure}

Fig. \ref{Per-user delay} presents the comparative results of the end-to-end latency of each device in the three schemes, where the transmission power of each mobile device is bounded by 24 dBm. In this simulation, since both communication and computation latency increase almost linearly with the number of devices, the end-to-end latency of each scheme has an approximately linear trend. From the figure, the local computing scheme always delivers the worst latency performance among three schemes because of the insufficient computational resource of mobile devices. Moreover, due to the limited communication resource, the end-to-end latency increases with the number of devices in all schemes. However, the proposed edge computing scheme always achieves the best latency performance since it can collaborate the cloud computing and edge computing. Specifically, for the case with 36 devices, the edge computing scheme can reduce about 41.44\% and 12.85\% latency as compared with local computing and cloud computing schemes, respectively. Note that in the simulation, we set the computation capacity of the cloud server almost 10 times of the edge server as a typical example. With the increase of the cloud computation capacity, the performance gap between the cloud computing scheme and the edge computing scheme will become smaller.

Fig. \ref{Total-user energy} depicts the comparative results of the total energy consumption of all devices with the number of devices, where the maximum delay tolerance of each AR application is 450 ms. From this figure, the cloud computing scheme can always achieve a better energy efficiency than the local computing scheme because of the extra energy consumption for tracking and mapping at local devices. Nevertheless, the proposed edge computing scheme always achieves the highest energy efficiency among all schemes. In the scenario with 36 devices, our proposed scheme can save about 73.71\% and 65.34\% energy consumption as compared with local computing and cloud computing schemes, respectively. Specifically, the overall energy consumption of each scheme would grow rapidly when the number of devices becomes large. The reason is that the communication energy becomes dominant with large number of mobile devices, which increases exponentially with the number of devices.

To further show the performance advantage on energy saving of the proposed edge computing scheme, we depict the total energy consumption of all mobile devices with different delay tolerances in Fig. \ref{Total_Energy_Delay_tolerance}, where the number of devices is set to be 30. From the figure, the mobile energy consumption of each scheme decreases with the maximum delay tolerance. The reason is that, more computational resource should be utilized to assist the task processing to meet the stricter delay requirement, leading to the higher mobile energy consumption. Once again, the superiority of the edge computing scheme is demonstrated by this test.

Overall, our tests demonstrate that the MEC-based AR framework is more efficient than other two conventional schemes as a result of the collaboration between cloud computing and edge computing. Both the processing latency and the energy consumption can be significantly reduced, leading to better QoE, such as high image resolution, long battery autonomy, convenient portability, and real-time performance for AR subscribers.

\section{Conclusion}
This article presents a novel MEC-based computation framework for current AR applications to optimize the energy efficiency and processing delay. A hierarchical computation architecture is first developed, which is composed of three layers: the user layer, the edge layer, and the cloud layer. By seamlessly integrating the communication, computation, and control functions at the edge layer and taking full advantages of both edge computing and cloud computing, computation-intensive tasks of AR applications can be intelligently offloaded to both edge and cloud servers for collaborative computation. In accordance with this architecture, we then develop a novel mechanism to simultaneously support multiple AR applications of different mobile subscribers. Three key technologies are also discussed, which can cooperate to further reduce the processing latency and energy consumption of mobile AR devices. Our proposed architecture and mechanism are finally tested by simulation results, which  demonstrate the substantial performance improvement of our proposals over the existing schemes.

\end{document}